# Enhancing Channel Assignment Performance in Wireless Mesh Networks Through Interference Mitigation Functions

M Pavan Kumar Reddy, Srikant Manas Kala and Bheemarjuna Reddy Tamma
Department of Computer Science and Engineering
Indian Institute of Technology Hyderabad, India Email
: [cs12b1025, cs12m1012, tbr]@iith.ac.in

*Abstract*—The notion of Total Interference Degree (TID) is traditionally used to estimate the intensity of prevalent interference in a Multi-Radio Multi-Channel Wireless Mesh Network (MRMC WMN). Numerous Channel Assignment (CA) approaches, link- scheduling algorithms and routing schemes have been proposed for WMNs which rely entirely on the concept of TID estimates. They focus on minimizing TID to create a minimal interference scenario for the network. In our prior works [1] and [2], we have questioned the efficacy of TID estimate and then proposed two reliable interference estimation metrics viz., Channel Distribu- tion Across Links Cost ($CDAL_{cost}$) and Cumulative X-Link-Set Weight ($CXLS_{wt}$). In this work, we assess the ability of these interference estimation metrics to replace TID as the interference-minimizing factor in a CA scheme implemented on a grid MRMC WMN. We carry out a comprehensive evaluation on ns-3 and then conclude from the results that the performance of the network increases by 10-15% when the CA scheme uses $CXLS_{wt}$ as the underlying Interference Mitigation Function (IMF) when compared with CA using TID as IMF. We also confirm that $CDAL_{cost}$ is not a better IMF than TID and $CXLS_{wt}$.

## I. INTRODUCTION

Multi-hop transmissions are a primary characteristic of Wireless Mesh Networks (WMNs), especially the class of Multi-Radio Multi-Channel (MRMC) WMNs. The broadcast nature of wireless transmissions gives rise to transmission conflicts, when data is being transmitted on identical or overlapping channels by radios that are in close proximity. The detrimental impact of prevalent interference caused by transmission conflicts erodes the network capacity of MRMC WMNs and their ability to transfer data seamlessly with minimal latency [3]. Relaying the transmitted data over several hops to deliver it to its destination further exacerbates the adverse impact of interference, as authors demonstrate in [4], by determining the upper-bound of the realizable aggregate throughput in a WMN. Impact of interference on data traffic in a WMN is quite detrimental as it leads to high packet loss and increased link-layer delays, which in turn marginalize the Quality of Service (QoS) offered by the wireless network.

Interference mitigation techniques in WMNs include channel assignment (CA) to radios, link-scheduling, routing and beam-forming through directional antennas. CA problem is an NP-hard problem which remains centric to interference mitigation in WMNs. In this work, we classify and analyze the interference mitigation mechanisms of various interference-aware CA schemes. Further, we determine the ideal interfer- ence mitigation function that optimizes WMN performance.

## II. INTERFERENCE MITIGATION FUNCTIONS

We now introduce the concept of Interference Mitigation Function (IMF). We define IMF as a mechanism which involves an Interference Estimation Metric (IEM) to assess the intensity of radio conflicts in the WMN at that particular instant, and a subsequent intelligent action to minimize the interference levels. An IMF is often iteratively invoked in a CA scheme, until no further reduction in levels of interference is recorded. In this section we discuss some important IMFs from the literature.

### A. Total Interference Degree (TID)

Most interference-aware channel assignments rely on TID which serves as a theoretical measure of interference prevalent in a wireless network. TID equals the number of conflict- links in a wireless network. A conflict-link signifies a potential link conflict that may adversely impact a particular link transmission in a wireless network. The number of conflict- links of a given wireless link is also called its Interference Degree (ID), and the sum of IDs of all the wireless links gives the TID. TID is often directly employed as an IMF in CA schemes such as MIS in [5] and CCA in [6], where channels are assigned in a prudent fashion so as to minimize TID. TID minimization depends on appropriate node selection that can be carried out through graph theoretic concepts such as Breadth First Search [7], neighbor partitioning scheme [8] or by a combination of algorithmic tools such as Tabu Search and Genetic Algorithms [9]. At times, the use of TID may not be explicit but the concept is inherent if CA schemes make use of conflict graphs, potential communication graph, potential collision domain etc., as has been demonstrated in [10].

### B. Channel Distribution Across Links ($CDAL_{cost}$)

$CDAL_{cost}$ [1] is an IEM inspired by the notion of statistical evenness of channel allocation, which hypothesizes that a

balanced distribution of channels among radios will offer a high performance CA. This metric adopts a probabilistic approach towards selection of links, wherein the temporal factors are taken into consideration by randomly choosing a link if two or more links are present between any two nodes in a network. It is assumed that the nodes may communicate on either link at any time and both the links are selected for transmission with equal probability. Thus, first the number of links operating on each available channel is determined. As a probabilistic approach is used for link selection, the link-count for a channel may be fractional. $CDAL_{cost}$ is then determined by finding the standard deviation of the link-counts. Since the guiding principle is that a proportionate channel distribution is always prudent, the closer the value of $CDAL_{cost}$ to 0, the more balanced is the channel allocation and better is the CA.

C. Cumulative X-Link-Set Weight($CXLS_{wt}$)

$CXLS_{wt}$ is a recently proposed IEM [2] which considers both the spatial and statistical aspects of interference in a wireless network while assessing its intensity. In this technique, the Transmission to Interference range (T:I) of a node is denoted as 1:X. The value of X determines the impact of interference caused by a wireless link on other links. The elemental unit of this algorithm is X-Link Set (XLS) which is any set of X consecutive links in the WMN topology. The $CXLS_{wt}$ algorithm finds all possible XLSs in the network. A particular weight is assigned to each XLS, where the weight is number of X links operating on non- interfering channels. All possible channel allocations are taken into consideration and thus weight for each case is determined. The final weight of the XLS is the mean of all such weights. For each possible scenario, the weight of XLS is assigned 0 (minimum) if all links operate on a common channel and the weight is X (maximum) in case all links operate on different channels. In other cases, the weight of XLS is the number of links functioning on non-interfering channels. The $CXLS_{wt}$ of WMN is sum of weights of all possible XLSs. Since a higher weight is assigned to an XLS with fewer conflicting links, higher the $CXLS_{wt}$ of the WMN better is the expected performance of the CA.

Both $CDAL_{cost}$ and $CXLS_{wt}$ are recent IEMs, and have been shown to be more accurate than TID in predicting the CA performance in [2] and [1]. However, no interference aware CA schemes have yet used them in IMFs. A plausible reason could be the absence of any research work which demonstrates that the twin IEMs are not just better estimates than TID, but can also be used as effective IMFs in CA design. In this work, we focus on bridging that gap and carry out a comprehensive evaluation of $CDAL_{cost}$ and $CXLS_{wt}$ vís-à-vís TID as prospective IMFs for CA schemes.

### III. CA Classification Based on IMF Module Placement

Since this work focuses on determining the best IMF for interference aware CA schemes, it is pertinent to identify and study the implementation of IMF modules within the CA mechanisms. We study numerous CA schemes and propose a four fold classification of CAs based on the IMF module placement in the CA architecture. This classification also tracks the development of CA formulation techniques and tracks their evolution from a simplistic linear brute-force approach to a complex layered hybrid approach. We also develop and present the generic high-level schemas of CA architecture of each CA class.

A. Bruteforce Interference Optimization (BIO)

These CAs carry out a rudimentary, computationally intensive and enumerative interference optimization by considering all possible scenarios of channel allocation to radios in a WMN, computing interference estimate of each and determin- ing the minimal interference scenario. For a WMN with $n$ nodes, $m$ radios per node and $c$ available channels, then the number of possible ways to allocate channels to the radios are $c_{(n*m)}$.

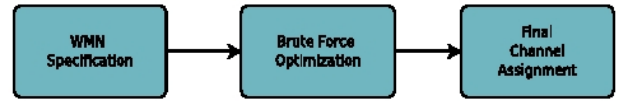
Fig. 1: Bruteforce Interference Optimization

In Figure 1, we present the architecture of the BIO class CAs, which has a linear flow and is fairly simple. BIO implementations were in vogue in the nascent phase of research in WMNs, and is now rarely employed owing to its excessive computational overhead. Secondly, most MRMC WMN nodes support dynamic switching to most efficient channels which needs to happen in the order of $\mu$seconds. Nevertheless, we consider BIO CAs in our classification because of their optimality i.e., as a benchmark against which performance of other CA classes can be assessed. A well-known BIO class CA is the list-coloring CA proposed in [11] which employs a brute-force mechanism to determine an optimal solution. Authors in [12] also highlight the use of brute-force techniques to CA problems.

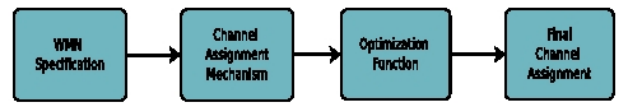
Fig. 2: Post Channel Allocation Interference Optimization

B. Post Channel Allocation Interference Optimization (PIO)

These CAs represent a slight enhancement over the BIO class, although the CA architecture remains linear as is evident from the schema presented in Figure 2. These CAs exploit heuristic techniques, graph theoretic approaches, genetic algorithms, etc. to carry out a preliminary channel allocation, at a significantly reduced computational cost. The initial CA is then processed further through one or more iterations of an IMF to obtain a more efficient and prudent final CA. Thus, the characteristic feature of CAs of this class is a 2-step mechanism. This can be seen in the load-aware CA proposed in [8] which employs neighbor partitioning, considering number of interfaces and available channels in the first step. To achieve minimal interference, in the second step, it assigns a channel to each node which is least used by its neighboring nodes. Similarly in [10], authors propose a centralized technique that constructs the maximization of throughput as a single commodity flow problem. First, the links of each node are grouped based on their local flows. The impact of interference is then alleviated by sorting the groups assigning channels to groups in the sorted order. This two-step mechanism can be seen in a plethora of CA schemes of the PIO class.

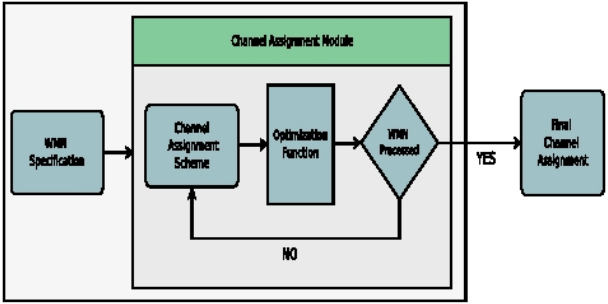

Fig. 3: CA Kernel Interference Optimization

### C. Kernel Optimization (KO)

This class is the next phase in CA design evolution, wherein the schema is complex and the IMF is embedded into the CA allocation mechanism, to form a Channel Assignment Module (CAM). Usually, these CAs employ a feedback mechanism to perform multiple iterations of CAM, until the CA can no longer be optimized. The CA architecture is presented in Figure 3. CAs of this class are very efficient and exhibit excellent network performance. There are no clear functional divisions or steps in the CA mechanism and multiple iterations are the primary characteristic. We list three well known CAs that rightfully belong to this class. A genetic algorithm based CA proposed in [13] strives to reduce the total interfering traffic load over the WMN by adopting random selection, crossover and mutation in multiple iterations. Authors present a Genetic Taboo Search based CA in [9] which uses Maximal Independent Sets and assigns channels to the links in a proba- bilistic way. It enhances the channel assignment performance over a number of iterations and each such iteration is a KO step.

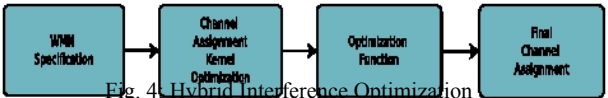

Fig. 4: Hybrid Interference Optimization

### D. Hybrid Optimization (HO)

Hybrid Optimization CAs are the climax of CA evolution. They are the most sophisticated and offer best performance in terms of network capacity, end-to-end latency and network resilience. Their design invariably involves a KO step and may exhibit the characteristics of one or more of the the other two phases. In addition, they may also have some special characteristics of their own. HO class CAs can be represented by the schema depicted in Figure 4. A suitable example is the Adap- tive Dynamic Channel Allocation (ADCA) scheme proposed in [14] which takes throughput and delay into consideration for channel assignment, and works well for both static and dynamic interfaces. Very similar to a KO class CA, the CAM is invoked several times to optimize performance. But it further optimizes the throughput from end-users to gateways for static nodes and determines the least-contention common channel in an on-demand fashion for dynamic nodes. Radio Co-location Aware Elevated Interference Zone Mitigation CA(RCA-EIZM- CA) [15] is an another CA which adopts the HO technique where the centralized principle of the algorithm is based on the fact that the prevailing interference in the network fluctuates creating localized pockets of unusually high interference levels and those localized pockets are called Elevated Interference Zones (EIZs). The CA proceeds by identifying the EIZs and assigning them channels so that the network performance increases at each KO step. The final step of this CA involves Radio Co-location Interference (RCI) mitigation by removing the co-located radios with same channel, adopting a step from PIO phase. EIZM-CA exhibits a remarkable feature. It is designed in such a way that its constituent functions can be disassembled and the TID based IMF can be replaced by any other IMF. This feature makes EIZM-CA a suitable candidate for our experimental study. We exploit its modular architecture to replace the IEM in its IMF, with $CDAL_{cost}$ and $CXLS_{wt}$ to test their effectiveness.

## IV. SIMULATIONS, RESULTS AND ANALYSIS

We now test the correctness of the proposed work of using $CDAL_{cost}$ and $CXLS_{wt}$ as the optimization functions in reducing interference. We run simulations on 802.11g ns-3 environment. Since EIZM-CA adopts the HO classification, we use this channel assignment as the reference and we develop the PIO and KO versions of EIZM-CA for analysis purpose. The primary reason for adopting EIZM-CA is that it not only factors the spatial characteristics of interference but also considers the components in the statistical dimension. EIZM-CA also correlates interference on the link to Signal- to-Interference-plus-Noise Ratio (SINR) of the link directly with the aid of EIZs. We generate the KO version of the CA by eliminating the final step which involves RCI mitigation and then generate the PIO version from the KO version CA by restricting the number of iterative steps to one. Although EIZM-CA uses TID as the elementary IMF, its design helps us to generate CAs with other IMFs. In order to test how $CXLS_{wt}$ and $CDAL_{cost}$ serve as IMFs, we generate new versions of EIZM-CA with $CXLS_{wt}$ and $CDAL_{cost}$ as the underlying IMF which will be mentioned as $HO_{CXLS}$ and $HO_{CDAL}$. $HO_{TID}$ is the CA scheme which uses TID as IMF. We do the same for the other two phases, PIO and KO , which will be referred as $PIO_{CXLS}$, $PIO_{CDAL}$, $PIO_{TID}$, $KO_{CXLS}$, $KO_{CDAL}$ and $KO_{TID}$.

### A. Simulation Parameters

We perform rigorous simulations in 802.11g ns-3 [16] environment to track the performance of different EIZM CAs obtained as a result of using TID, $CDAL_{cost}$ and $CXLS_{wt}$ as the optimization functions in a $5 \cdot 5$ grid WMN. The reason for selecting a WMN of grid layout is because of its advantages when compared with any random grid in terms of coverage area, total network capacity, capacity to withstand heavy traffic and most importantly it is a perfect scenario for evaluating the performance of any CA [17]. The parameters for the simulations are presented in Table I. There will be 10 concurrent 4-hop flows, five horizontal and five vertical, starting from the first node of a row or column and ending at the last node of that particular row or column. In this test scenario, all the nodes will be exclusively used for data trans- mission so that we can thoroughly inspect the performance of CA. Each of these flows transmits a file of size 5 MB from source node to destination node. The simulations are handled with Transmission Control Protocol (TCP) and User Datagram Protocol (UDP) as the transport layer protocols in different sets of experiments. These protocols are implemented through the inbuilt BulkSendApplication and UdpClientServer models in ns-3. Throughput and SINR values are retrieved through the TCP simulations, meanwhile Packet Loss Ratio (PLR) and Mean Delay (MD) are claimed from UDP simulations. Each simulation is carried out for five seeds, with a simulation time of 2000 seconds. Constant Rate WifiManager is used as the underlying rate control algorithm and the simulations were run for both 9 and 54 Mbps PHY data rates.

TABLE I: NS-3 Simulation Parameters

| Parameter | Value |
| --- | --- |
| IEEE Protocol Standard | 802.11g |
| Number of Orthogonal Channels Utilized | 3 (2.4 GHz) |
| Datafile Size | 5 MB |
| 802.11g PHY Datarate | 9/54 Mbps |
| TCP ns-3 model | BulkSendApplication |
| UDP ns-3 model | UdpClientServer |
| TCP Maximum Segment Size | 1 KB |
| UDP Packet Size | 1024 Bytes |
| MAC Fragmentation Threshold | 2200 Bytes |
| TCP RTS/CTS handshake | Enabled |
| UDP RTS/CTS handshake | Disabled |
| Routing Protocol | OLSR |
| Rate Control | Constant Rate |
| UDP Inter Packet Interval | 0.05 seconds |

### B. Results and Analysis

Meticulous simulations are run and the values of the performance metrics viz., Network Throughput (Mbps), PLR (% of packets lost) , MD($\mu$seconds) and SINR (db) are measured. The results are shown in Figures 5, 6, 8, 7 and Tables II, III and IV.

From the Figures 5, 6, 7 and 8, it can be established that the BIO in CA scheme with the underlying IMF as $CXLS_{wt}$ performs better in case of all the four performance metrics. It can also be concluded that BIO in CA scheme where the underlying IMF is $CDAL_{cost}$ does not perform better than BIO in CA schemes with TID and $CXLS_{wt}$ as IMFs. This goes to prove that $CXLS_{wt}$ is a better IMF when compared with TID and $CDAL_{cost}$. With this result as a basis, we can imply that for any given CA scheme using IMF, the CA scheme will result in a better channel assignment when the IMF is $CXLS_{wt}$ than when IMF is $CDAL_{cost}$ or TID.

To test the efficacy of $CDAL_{cost}$ and $CXLS_{wt}$ as the IMFs, we generated the PIO, KO, HY versions of EIZM-CA with all the three IEMs viz., TID, $CDAL_{cost}$ and $CXLS_{wt}$ as the underlying IMFs. Tables II, III and IV show the results of multiple versions of EIZM-CA using different IMFs. The results convey that as we move from the simple and linear flow PIO version to complex and iterative structured HO, the performance of the CA gets better at each step. There is a fair amount of improvement in case of PIO to KO transition but the improvement is significant in case of KO to HO. The increase of SINR values from KO to HO is very noteworthy implying that interference levels have significantly decreased. These results validate that the performance of the CA scheme will vary depending upon the application of IMF modules within the CA mechanisms irrespective of the IEM. The results of HO and BIO versions are also very comparable proving that as we move to the better versions of CA, we arrive at the optimal BIO version of CA.

We can infer form the results in Table II that $CXLS_{wt}$ works as the better IEM in all the performance metrics outrunning $CDAL_{cost}$ and TID in PIO at both 54 & 9 Mbps PHY data rates. In case of KO, $CXLS_{wt}$ performs better except for the PLR metric in 54 Mbps data rate where the $CDAL_{cost}$ outruns $CXLS_{wt}$ by a slight margin. In case of 9 Mbps, a similar trend where $CDAL_{cost}$ performs better by a negligible margin can be observed in case of PLR and SINR. In HO also, $CXLS_{wt}$ performs better in all the performance metrics when compared with $CDAL_{cost}$ and TID where the underlying data rate is 54 Mbps. For the 9 Mbps simulations, in case of PLR, MD and SINR, $CDAL_{cost}$, TID and both perform better than $CXLS_{wt}$ with a very less verge. Since the PIO is simple and linear, the margin of improvement for $CXLS_{wt}$ when compared with TID and $CDAL_{cost}$ is very less. As the complexity increases, the margin of improvement increases as we can clearly observe in the cases of KO and HO. It can be incurred from the results that in case of throughput, $CXLS_{wt}$ works perfectly in both 9 and 54 Mbps PHY data rates. Meanwhile in case of the other three performance metrics PLR, MD and SINR, the results are in accordance with the Brute Force CAs in 54 Mbps indicating that $CXLS_{wt}$ is a better IMF than TID and $CDAL_{cost}$. In case of 9 Mbps simulations for the other three performance metrics, either TID or $CDAL_{cost}$ outruns $CXLS_{wt}$ by a very

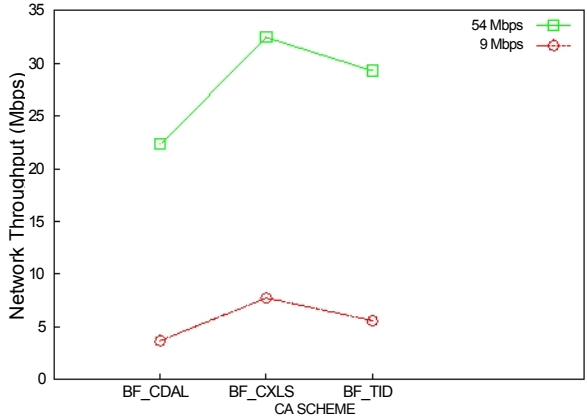
Fig. 5: Throughput in BF CAs

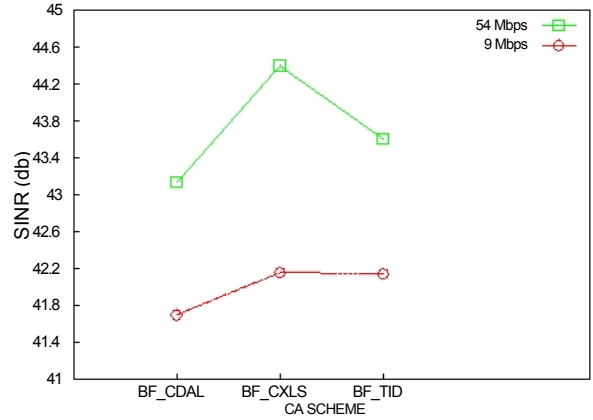
Fig. 6: SINR in BF CAs

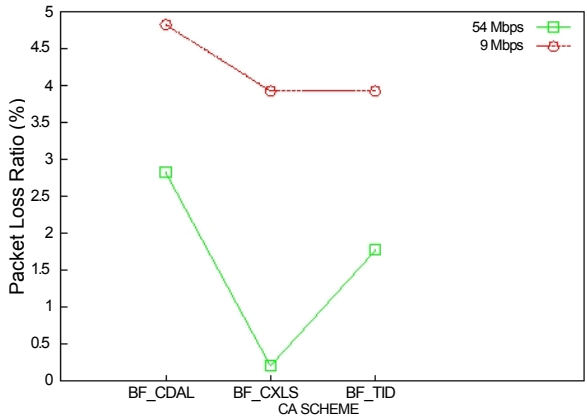
Fig. 7: Packet Loss Ratio in BF CAs

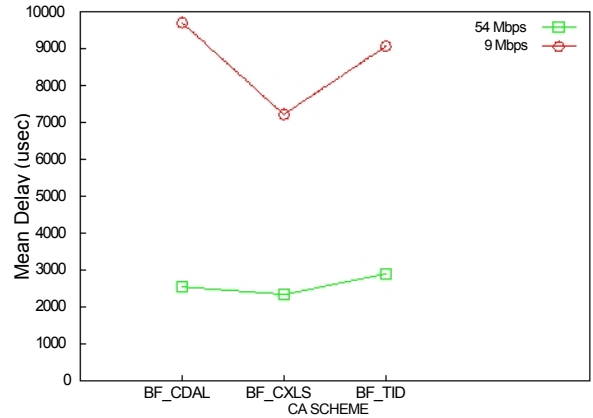
Fig. 8: Mean Delay in BF CAs

TABLE II: Performance Evaluation of CAs in PIO using different IMFs

| Performance Metric | CHANNEL ASSIGNMENT SCHEME | | | | | |
| --- | --- | --- | --- | --- | --- | --- |
| | $PIO_{CDAL}$ | | $PIO_{CXLS}$ | | $PIO_{TID}$ | |
| | 9 Mbps | 54 Mbps | 9 Mbps | 54 Mbps | 9 Mbps | 54 Mbps |
| Throughput(Mbps) | 2.59 | 15.32 | 4.37 | 22.47 | 4.08 | 20.24 |
| PLR (%) | 13.91 | 16.8 | 0.6 | 3.94 | 3.19 | 7.29 |
| MD($\mu s$) | 13432 | 2981 | 7944 | 1764 | 11756 | 2863 |
| SINR (db) | 9.98 | 9.97 | 10.05 | 12.65 | 10.03 | 12.31 |

less margin.

The results prove that $CXLS_{nt}$ is not only a better IEM [2], but also functions as a better IMF when compared with TID and $CDAL_{cost}$. The results also conclude that $CDAL_{cost}$ does not perform better when compared with TID and $CXLS_{nt}$ as an IMF. So it can be incurred that though $CDAL_{cost}$ is a better IEM [1] than TID, it fails as IMF. This might be because of the fact that $CDAL_{cost}$ considers only the statistical features of interference by assuming that even distribution of channels among the radios will result in a better CA, which is not the case in the real world scenarios entirely overlooking the spatial and temporal aspects of interference. The reason for $CXLS_{nt}$ performing better than the other IEMs is that it takes both the statistical and spatial features into consideration which the other IEMs does not. Similar trends are observed in other WMN configurations as well, where we vary one or more WMN specifications viz., the size of a grid or a random WMN, the number of available channels or the number of radios per node.

## V. CONCLUSIONS

In order to determine the IMF, which results in better CA for a given channel assignment scheme, we classified CAs into four categories based on the implementation of the IMF module. The classification varies from a simple and linear brute force approach to a complex hybrid approach, wherein the Brute Force CAs serve as the optimal benchmarks against which we evaluate the other CA classes. Having classified the CAs and evaluated the performance of various IMFs in all the CA classes, we now make a few valid conclusions. As the complexity of the approach increases from PIO to HO, the performance of CA increases, proving that irrespective of the IEM employed CA performance is intricately linked to invocation of IMF at appropriate flow-points in the CA design. The performance of Hybrid Optimization CAs is very close to optimal Brute Force CAs, which indicates the convergence towards optimality as CA architecture grows more complex. $CXLS_{wt}$ outperforms TID and $CDAL_{cost}$ as an IMF in most cases, owing to its spatio-statistical character. Although $CDAL_{cost}$ is a better IEM when compared to TID, results demonstrate that it does not perform better as an IMF, as it takes only statistical features of interference into consideration.

TABLE III: Performance Evaluation of CAs in KO using different IMFs

| Performance Metric | CHANNEL ASSIGNMENT SCHEME | | | | | |
|---|---|---|---|---|---|---|
| | $KO_{CDAL}$ | | $KO_{CXLS}$ | | $KO_{TID}$ | |
| | 9 Mbps | 54 Mbps | 9 Mbps | 54 Mbps | 9 Mbps | 54 Mbps |
| Throughput (Mbps) | 4.43 | 19.81 | 7.26 | 23.98 | 7.25 | 22.76 |
| PLR (%) | 1.32 | 3.14 | 3.94 | 6.26 | 6.22 | 13.82 |
| MD ($\mu s$) | 12892 | 2757 | 9719 | 1651 | 9391 | 1883 |
| SINR (db) | 10.09 | 11.09 | 10.03 | 11.64 | 10.02 | 10.86 |

TABLE IV: Performance Evaluation of CAs in HO using different IMFs

| Performance Metric | CHANNEL ASSIGNMENT SCHEME | | | | | |
|---|---|---|---|---|---|---|
| | $HO_{CDAL}$ | | $HO_{CXLS}$ | | $HO_{TID}$ | |
| | 9 Mbps | 54 Mbps | 9 Mbps | 54 Mbps | 9 Mbps | 54 Mbps |
| Throughput (Mbps) | 4.39 | 21.87 | 7.15 | 27.44 | 6.76 | 25.51 |
| PLR (%) | 4.17 | 0.87 | 2.38 | 0.83 | 2.20 | 1.02 |
| MD ($\mu s$) | 7236 | 2520 | 8315 | 2387 | 9351 | 2440 |
| SINR (db) | 42.52 | 44.02 | 43.34 | 44.04 | 43.42 | 43.95 |